\documentclass[letterpaper,12pt]{article}

\usepackage{graphicx} 
\usepackage{amsmath}
\usepackage{multirow}

\begin{document}

\begin{center}{\Large
In Situ Evaluation of Density, Viscosity and Thickness of Adsorbed Soft Layers by Combined Surface Acoustic Wave and Surface Plasmon Resonance.}

{Laurent A. Francis $^\star$$^{1,2}$, Jean-Michel Friedt$^3$, Cheng Zhou$^2$, and\\ Patrick Bertrand$^1$} \\
$^1$ Unit\'e de Physico-Chimie et de Physique des Mat\'eriaux, \\ Universit\'e
catholique de Louvain, 1 Croix du Sud, B-1348 Louvain-la-Neuve, BELGIUM\\
$^2$ Microsystems, Components and Packaging Division, IMEC, 75 Kapeldreef, B-3001 Leuven, BELGIUM \\
$^3$ FEMTO-ST/LPMO, 32 Avenue de l'Observatoire, 25044 Besan\c con, FRANCE\\
\end{center}
\vspace{2cm}
Corresponding author: Laurent FRANCIS, IMEC, MCP-TOP, Kapeldreef 75, B-3001 Leuven, Belgium, phone: +32-16-288564, fax: +32-16-288500, e-mail: francisl@ieee.org

\newpage
\vfill {\Large Abstract} \vspace*{0.5cm}\\

We show the theoretical and experimental combination of acoustic and optical methods for the in situ quantitative evaluation of the density, the viscosity and the thickness of soft layers adsorbed on chemically tailored metal surfaces. For the highest sensitivity and an operation in liquids, a Love mode surface acoustic wave (SAW) sensor with a hydrophobized gold coated sensing area is the acoustic method, while surface plasmon resonance (SPR) on the same gold surface as the optical method is monitored simultaneously in a single set-up for the real-time and label-free measurement of the parameters of adsorbed soft layers, which means for layers with a predominant viscous behavior. A general mathematical modeling in equivalent viscoelastic transmission lines is presented to determine the correlation between experimental SAW signal shifts and the waveguide structure including the presence of the adsorbed layer and the supporting liquid from which it segregates. A methodology is presented to identify from SAW and SPR simulations the parameters representatives of the soft layer. During the absorption of a soft layer, thickness or viscosity changes are observed in the experimental ratio of the SAW signal attenuation to the SAW signal phase and are correlated with the theoretical model. As application example, the simulation method is applied to study the thermal behavior of physisorbed PNIPAAm, a polymer whose conformation is sensitive to temperature, under a cycling variation of temperature between $20$ and $40\ \mathrm{^\circ C}$. Under the assumption of the bulk density and the bulk refractive index of PNIPAAm, thickness and viscosity of the film are obtained from simulations; the viscosity is correlated to the solvent content of the physisorbed layer.
\newpage





\section{Introduction}
Sorption processes at the solid/liquid interface by which (bio)molecules bound to material surfaces are of interest for biosensors, biomaterials, materials characterization and surface science. Understanding the three-dimensional organization (including density, water content and thickness) of the resulting sorbed film and its evolution during the process is crucial for many applications in these domains. For biosensors, more specifically, there is a need to monitor the response in real-time and to be able to distinguish contributions coming from the dry sorbed mass, which is the physical criterion for estimating sensitivity, and those that should be attributed to effects intimately associated to the film organization, like sorbent-bound water and viscosity for example. While a wide variety of methods can qualitatively detect the formation of sorbed films, almost none of them, alone, is quantitative and able to reveal the film organization. Furthermore, only a few techniques can monitor the sorption process in real-time. Scanning probe microscopies might fulfill all these requirements, mainly for submonolayers sorbed on ultra smooth surfaces \cite{Keller00}. Neutron reflectivity \cite{Lu00}, XPS \cite{Plunkett03}, mass spectroscopy \cite{Hanley99} and radiolabeling \cite{Dufrene99} are quantitative techniques able to directly measure the dry sorbed amount, nevertheless compatible to very different film organizations. Of all the direct detection (i.e. label-less) techniques, we have identified acoustic and optical methods as being the only ones fulfilling two fundamental criteria of our measurements: time resolved and in situ (liquid phase) measurement of the physical properties of the adsorbed layer.\\

Various methods of direct detection of biochemical layers have been developed, either based on the disturbance of an acoustic wave \cite{Cavic99} (quartz crystal microbalance \cite{rodahl02} -- QCM -- and surface acoustic wave devices \cite{gizeli2,mchale,jakoby,jakoby2,kourosh,gizeli3,harding} -- SAW) or of an evanescent electromagnetic wave (optical waveguide sensors \cite{Cush93,Voross02} and surface plasmon resonance \cite{Liedberg83,gizeli1} -- SPR). While each one of these transducers provides reliable qualitative curves during the adsorption of a soft layer on their functionalized surface, such as proteins for example, extraction of quantitative physical parameters such as optical index, density, viscosity or water content requires modeling of the adsorbed layers \cite{Hook01}. The modeling includes multiple parameters which must be identified simultaneously: hence the need for the combination of acoustic and optical detection methods in a single instrument \cite{Laschitsch00,Bailey02,Bund03,Plunkett203}.\\

We here use a combination of Love mode surface acoustic wave (SAW) device and surface plasmon resonance (SPR) and present a methodology to extract out of experimental data and simulations a quantitative value for the density, the viscosity and the thickness of adsorbed soft layers from a liquid phase. The viscosity and the density will be further correlated to the water content in the adsorbed layer. As a showcase of the performances of our methodology, the detection of the physisorption of poly(N-isopropyl acrylamide) -- PNIPAAm -- is presented. PNIPAAm is a thermally sensitive polymer with a conformal transition investigated previously by SPR and QCM techniques \cite{Balamurugan03,Plunkett203}. The chemical structure of PNIPAAm is reported in Fig.~\ref{fig:pnipaam}. Our approach is presented as a sequel to the study by Plunkett {\it{et al.}} \cite{Plunkett203} where the properties of PNIPAAm physisorbtion to a hydrophobic surface was monitored by a combination of SPR and QCM, and the subsequent temperature-dependent conformation transition was monitored by QCM alone by measuring both the resonance frequency and the quality factor of the resonator. In the present study, Love mode SAW devices were chosen for a high mass sensitivity and a compatibility with measurements in liquid media \cite{Wang94,Du96,Gizeli97,McHale02}. Being based on the propagation of a shear-horizontal acoustic wave, their interaction with the surrounding liquid is minimal although the bulk properties of the solvent influence the acoustic wave propagation and must be considered in the analysis of the sensing device \cite{Jakoby98,Herrmann99} and a model for the analysis of the structure, including the adsorbed layer and the liquid, is presented in the theoretical section (section \ref{theory}). While both SAW and SPR sensors display theoretically similar sensitivities and detection limits, acoustic sensors are typically sensitive to temperature fluctuations unless optimized to achieve a zero temperature coefficient of frequency (TCF). In a normal operation, care must be taken to properly control the environment around the instrument to keep a stable temperature. However, in the present case, the temperature will be voluntary modified and monitored to modify the characteristics of the adsorbed PNIPAAm. This specific setting requires a preliminary calibration curve of the acoustic response as a function of the temperature. The analysis of the layer will assume that a certain amount of the solvent -- water in the present case -- is contained in the layer. A schematic representation of the experimental combined set-up used for this study is given in Fig.~\ref{fig:sawspr1}.

\section{Theory}\label{theory}

{\noindent{\bf{Love mode surface acoustic waveguide.}} Acoustic devices are sensitive to mechanical perturbations, temperature changes and electrical perturbations such as pH changes, conductivity and dielectric permittivity of added materials. If the sensing area is electrically shielded by a grounded conducting layer, these electrical effects can be neglected, letting the device be sensitive to mainly mechanical perturbation. For the model, we will assume that a composite material segregates from the bulk solution on the waveguide surface as an hypothetical perfectly flat layer of given characteristics to be determined. Acoustic waves with a shear-horizontal polarization are best suited for sensing applications in liquids: the mechanical displacement at the surface probes within a certain volume the density, the shear stiffness and the dynamic viscosity of the composite material that has a certain thickness and is adsorbed on the surface, thus resulting in a mechanical interaction between the layer and the acoustic device. The Love mode acoustic structure supports the guided propagation of a shear-horizontal wave in a stack of guiding layers deposited atop a piezoelectric crystal substrate. For a large velocity difference between the guiding layers and the substrate, the acoustic energy is trapped closely to the sensing area and results in the highest sensitivity of all similar acoustic structures. The acoustic velocity and the attenuation of the acoustic wave in the stacked structure is electrically probed thanks to interdigitated electrodes patterned on the piezoelectric material. The phase at a given frequency within the bandpass range of the delay line and the insertion loss of the acoustic device measured electrically, for instance with a network analyzer, are representative of the actual propagation characteristics of the acoustic wave along the delay line and, in particular, along the sensing path.}\\

{\noindent{\bf{Equivalent transmission line model.}} It is known that the propagation of acoustic waves in guiding structures can be obtained by equivalent transmission lines \cite{Oliner65,Auld90:I}.} An implementation of this equivalent model is given here to analyze the propagation characteristics of the guided wave in the Love mode structure, including the presence of the surface adsorbed composite layer and the liquid from which this layer segregates. At first, each material is equivalently represented by a transmission line with a wavevector $\Vec{k}$ and a characteristic impedance $Z$ such that the shear stress $T(\Vec{r})$ and shear particle velocity $v(\Vec{r})$ fields at a given point $\Vec{r}$ in the material are given by the superposition of incident and reflected acoustic waves:
\begin{eqnarray*}
T(\Vec{r}) & = & T_{+} \ e^{-\Vec{k} \cdot \Vec{r}} + T_{-} \ e^{\Vec{k} \cdot \Vec{r}}\\
v(\Vec{r}) & = & \frac{T_{+}}{Z} \ e^{-\Vec{k} \cdot \Vec{r}} - \frac{T_{-}}{Z} \ e^{\Vec{k} \cdot \Vec{r}}
\end{eqnarray*}
where $T_{+}$ and $T_{-}$ are arbitrary values for the intensity of the incident and reflected waves. The equivalent representation of a viscoelastic layer in this model is given in Fig.~\ref{fig:rlc_model}, which consists in a series inductance $L$ shunted by a capacitance $C$ placed in series with a conductance $G$. The values of the equivalent parameters are given in mechanical terms of the layer, i.e. its density $\rho$, its shear stiffness $\mu$ and its dynamic viscosity $\eta$ by the following correspondence \cite{Auld90:I}
\begin{eqnarray*}
L &=& \rho \\
C &=& \frac{1}{\mu} \\
G &=& \frac{1}{\eta}
\end{eqnarray*}
from which we derive an analytical expression of the wavenumber $k=\left|\Vec{k}\right|$ and of the characteristic impedance $Z$:
\begin{eqnarray*}\label{eq:kandZ0}
k &=& \frac{\mathrm{i}\omega}{V} \\
Z &=& \frac{\mu+\mathrm{i}\omega \eta}{V}
\end{eqnarray*}
where $\omega = 2\pi f$ is the angular frequency, while $V$ is the complex acoustic velocity defined by
\begin{equation*}\label{eq:Vcomplex}
V = \sqrt{\frac{\mu+\mathrm{i}\omega \eta}{\rho}}.
\end{equation*}
For waveguides, the wavenumber $k$ is decomposed in a resonance wavenumber $k_z$ and a propagation wavenumber $k_x$ such that $k^2 = k_x^2+k_z^2$. The projection into both directions, propagation and resonance, is expressed in each layer by a \textit{complex coupling angle} $\varphi$ such that
\begin{eqnarray} \label{eq:kjx}
k_{jx} &=& k_{j}\sin{\varphi_j}\\ \label{eq:kjz}
k_{jz} &=& k_{j}\cos{\varphi_j}\\ \label{eq:Zjx}
Z_{jx} &=& Z_{j}\sin{\varphi_j}\\ \label{eq:Zjz}
Z_{jz} &=& Z_{j}\cos{\varphi_j}
\end{eqnarray}
where the subscript $j$ refers to a constituent layer. The {\it{transverse resonance principle}}\cite{Oliner65} is applied to determine the propagation wavenumber: according to this principle, the coupling angles are related between each others such that the propagation wavenumbers $k_{jx}$ are all equal. For a simpler notation, the $k_{jx}$ being all identical are noted $k_{x}$. \\

To determine the propagation characteristics in the waveguide structure, the last step is to determine the resonance wavenumber. To that purpose, an equivalent circuit of the physical structure represented in Fig.~\ref{fig:tlm_mod12} is built by connecting in series each equivalent transmission lines of the layers stacked in the direction of resonance, as depicted in Fig.~\ref{fig:tlm_mod11}. The series connection automatically solves the boundary conditions associated to the continuity of the mechanical stress and strain at materials interfaces. As a consequence, semi-infinite layers (the substrate $S$ and the fluid $F$) are terminated by a matching load that accounts for the endless side while the other layers (the guiding layers $G_j$ and the composite layers $C_j$) have finite dimensions, respectively $t_j$ and $h_j$. One must notice that for a clearer representation, only one guiding layer and one composite layer have been schematically depicted although the model is easily extended to a larger amount of layers. Finally, the resonance is found by measuring the impedance at a plane $\mathrm{T}$ located at an arbitrary position in the structure, for instance at the substrate--coating interface as sketched in Fig.~\ref{fig:tlm_mod11}: the resonance is found for values of the complex coupling angle that result in opposite impedances seen in both directions at the reference plane $\mathrm{T}$. These impedances are determined by applying a standard formula from the transmission line theory to calculate the input impedance of a transmission lines of characteristic impedance $Z_{char}$ and length $L$ loaded at its end with an impedance $Z_{load}$:
\begin{equation*}
Z=Z_{char}\ \frac{Z_{load}+ Z_{char}\tanh \left( k_zL \right)}{Z_{char}+ Z_{load}\tanh \left( k_zL \right)}.
\end{equation*}
A solution of the resonance wavenumber is translated to the propagation wavenumber thanks to Eqs.~\eqref{eq:kjx} and \eqref{eq:kjz}. The real and imaginary parts of a determined solution $k_{x}$ are related to the attenuation and the phase velocity of a guided mode, respectively. \\

{\noindent{\bf{Electrical signal.}}} For the analysis of the acoustic signal, variations in amplitude and in phase are monitored during modifications of the mechanical characteristics of the composite layer or in its thickness. Because of such variations, the wavenumber solution changes to $k'_{x}$. The induced variations of the acoustic signal are measured through the transducers and interpreted as an electric signal attenuation per unit of length $\Delta A$, in $\mathrm{dB/m}$, given by
\begin{equation*}\label{eq:att}
\Delta A = 20\log_{10}{\frac{\mathrm{Re}(k_{x})}{\mathrm{Re}(k'_{x})}}
\end{equation*}
and to an electrical delay phase shift per unit of length $\Delta \phi$, in $\mathrm{1/m}$, given by
\begin{equation*}\label{eq:phivar}
\Delta \phi = \mathrm{Im}(k_{x})- \mathrm{Im}(k'_x).
\end{equation*}
The quantities to compare with experimental values are obtained by multiplying the theoretical shifts by the length of the sensing path $D$.\\

{\noindent{\bf{Rigid and viscous interactions.}} The ratio of the amplitude shift to the phase shift is a significant indicator of rigid or viscous interactions between the layer and the vibrating surface and will be used to discriminate the relaxation period of the adlayer. For any added layer on the sensing surface, four parameters must be evaluated from the measurement (density, shear stiffness, viscosity and thickness) while the SAW device gives only two independent measurements linked to the real and the imaginary part of the acoustic wavevector measured electrically by the transducers (insertion loss and phase); therefore, an assumption must be made to solve some complexity of the analysis to evaluate the unknown parameters. To that purpose, we assume that the composite layer and the surrounding liquid are Newtonian fluids, meaning that the shear stiffness in both layers, $\mu_C$ and $\mu_F$, is discarded in regard to the viscous components $\omega \eta_C$ and $\omega \eta_F$. This working hypothesis has been previously described in the literature \cite{Saha03} through the {\it{relaxation time}} $\tau = \eta/\mu$, and a comparison of the relaxation time to the excitation frequency $\omega$ driving the mechanical motion of the waveguide surface. The product of the relaxation time by the frequency indicates if the adsorbed film moves in phase (rigid) or out of phase (viscous) with the acoustically driven surface. When the relaxation time is high, i.e. when $\omega \eta \gg \mu$, the film is viscous, meaning that acoustic loss is a predominant mechanism of energy dissipation in the film, which is also a mechanism driving an important attenuation of the acoustic wave as it propagates along the perturbed part. At the opposite, when the relaxation time is low and $\omega \eta \ll \mu$, the film is considered as rigid, involving an entrapment of the acoustic energy in the film which participates effectively in the guiding of the acoustic wave.}\\

For viscous films, it has been demonstrated \cite{McHale03} that the product of the density by the viscosity and the film thickness play a role in the modification of the acoustic velocity and the attenuation of the signal; for rigid films, only the density and the film thickness play a role. A discrimination between both effects can be obtained through the ratio of the SAW signal $\Delta A/\Delta \phi$. The mechanical behavior of the film with respect to the excitation frequency is determined by the part, real or imaginary, of the propagation wavevector $k_x$ that is predominant. For the operating frequencies of Love mode devices, typically comprised between 80 MHz and 430 MHz, a film is seen as a viscous layer till shear stiffness up to $10^5$ Pa. Above this threshold, the film supports the propagation of the guided Love mode rather than its dissipation, resulting in a low ratio $\Delta A/\Delta \phi$ and leading to the influence of the density of the film only rather than the product of the density by the viscosity. The value of the ratio is obtained  from simulations for each specific Love mode device.\\

{\noindent{\bf{Penetration depth and film thickness.}}} The penetration depth of shear waves in fluids $\delta$ is an indication of the thickness probed by the acoustic method. The exact value of $\delta$ for surface acoustic waveguides is determined by the real part of the wavevector in the viscous layer as determined by the transmission line equivalent model. Its value is close to the one obtained from the well-known relation \cite{Saha03}
\begin{equation} \label{eq:delta1}
\delta = \sqrt{\frac{2\eta}{\rho \omega}}
\end{equation}
from which it slightly differs due to the coupling angle $\varphi_C$ in the viscous layer. Because of a usual large velocity difference between the rigid guiding layers and the viscous layers, the real part of the coupling angles $\varphi_C$ and $\varphi_F$ are almost equal to zero and, therefore, the penetration depth can be identified with the expression given by Eq.~\eqref{eq:delta1}. The acoustic wave senses the viscous layer with an exponential decay extending to a thickness about $3\delta$, after which the acoustic field decays below 10\% of its initial value and is neglected. This effect has an impact on the ratio $\Delta A/\Delta \phi$ as plotted in Fig.~\ref{fig:hoverdelta} where the ratio is plotted versus the thickness of the viscous film normalized to the penetration depth as simulated for different possible values of $\eta_C$. The ratio stabilises for film thicknesses above $\sim 3 \delta$, which is about 150 nm at a typical acoustic frequency of 123.5 MHz for a layer with $\eta_C = 1\ \mathrm{cP}$ (same as water). When the layer is thin ($< 0.1\delta$), the ratio becomes similar to the one obtained with a rigid layer; consequently, a small experimental ratio is not sufficient to conclude that the probed layer behaves as a viscous or a rigid layer. On the other hand, a high ratio is an unquestionable indicator of a viscous interaction, where the hypothesis of a soft layer can be assumed ($\omega \eta_C \gg \mu_C$). The estimation of the shear stiffness of the adsorbed film is removed from the film analysis while the density, the viscosity and the thickness remain to be determined. These three values can not be estimated from the two information (insertion loss and phase) resulting from a single SAW measurement. Consequently, an additional information about any of these three values is requested with a measurement technique compatible with surface acoustic waves. A first possibility is a thickness estimate by AFM \cite{Choi02}. A second possibility is an optical measurement of the film such as ellipsometry or surface plasmon resonance. This latter possibility has been chosen and is described here after.\\

{\noindent{\bf{Combined technique SAW/SPR.}} We have previously demonstrated the integration of a SAW device with surface plasmon resonance (SPR) in a single instrument for the simultaneous and real-time measurement of rigid adsorbed layers \cite{Friedt04}. The SPR signal is a measurement of the coupling plasmon angle as a function of variations in refractive index $n$ and thickness $h$ of the adsorbed layer. A single SPR measurement is therefore not sufficient to determine independently these two parameters. Our previous analysis applied to rigid layers, therefore the SAW shift was assumed to be proportional to the density and thickness of the layer with respect to a mass sensitivity calibration of the SAW device by copper electrodeposition. In the rigid layer assumption, the acoustic phase difference combined to the SPR signal is sufficient to estimate independently the density and the thickness of the layer under the assumption of a linear relationship between the variation of the density of the layer and its refractive index variation. This assumption, which we will assume that still hold in the present case, considers that both the density and the refractive index of the composite layer are directly correlated and vary between the bulk values of the component segregating from the fluid on the sensing surface alone and the bulk values of the fluid itself as a function of the quantity of fluid contained in the composite layer. If we denote by the term $x$ the relative amount of fluid in the composite layer, meaning that $x$ varies between 0 and 1, then the density $\rho_C$ of the composite layer is given by
\begin{equation}\label{eq:ad:rhoC}
\rho_C = (1-x)\rho_L + x\rho_F
\end{equation}
and its refractive index $n_C$ by
\begin{equation}\label{eq:ad:nC}
n_C = (1-x)n_L + xn_F
\end{equation}
where the subscript $F$ refers to the semi-infinite fluid layer, $C$ to the composite surface adsorbed layer, and $L$ to the bulk characteristics of the component segregating from the fluid.
By eliminating the unknown term $x$ between Eqs.~\ref{eq:ad:rhoC} and \ref{eq:ad:nC}, a direct relation between density and refractive index is expressed under the form:
\begin{equation} \label{eq:x}
\frac{\rho_C - \rho_F}{\rho_L - \rho_F} = \frac{n_C-n_F}{n_L-n_F}.
\end{equation}}

{\noindent{\bf{Quantitative analysis.}} The methodology to obtain an estimate of the density, viscosity and thickness of the soft layer is based on the principle described here after. for a potential value of the film thickness, the acoustic attenuation $\Delta A$ and the acoustic phase shift $\Delta \phi$ correspond, each of them, to a set of values for the density and the viscosity as obtained from the model described above. Considered together, the two sets present a common point, thus resulting in a couple values $[\rho_C(h),\eta_C(h)]$ as shown in Fig.~\ref{fig:rhoeta_saw1} where the lines of expected amplitude and phase shifts are plotted at a given thickness for a given value of the amplitude and phase shifts versus density and viscosity. For a set of values for the thickness, a curve of possible values of viscosity and density are obtained. An unique solution $[h,\rho_C,\eta_C]$ is obtained thanks to the SPR data: for each value of thickness at a given SPR angle shift $\Delta \theta$ corresponds an unique value of the refractive index \cite{Grossel94} that is further related to the density of the layer thanks to Eq.~\ref{eq:x}, which assumes an identical evolution of the refractive index and of the density of the composite layer. The methodology is schematically depicted in the following diagram:
\begin{equation*}
\left.
\begin{array}{r}
\left.
\begin{array}{r}
\Delta \phi\\
\Delta A
\end{array} \right\} \rightarrow [\rho_C(h),\eta_C(h)] \\
\Delta \theta
\end{array} \right\} \stackrel{Eq.\ref{eq:x}}{\rightarrow} [h,\rho_C,\eta_C]
\end{equation*}
Our method has the unique characteristics to provide the thickness, the density and the viscosity of the adsorbed layer out of three independent experimental measurements (phase and insertion loss of the SAW and coupling angle of the SPR), under assumed values for the density $\rho_L$ and the refractive index $n_L$ of the component and under the further assumption of an entirely viscous interaction between the composite layer and the acoustic wave (i.e. $\omega \eta_C \gg \mu_C$).\\

\section{Experimental section}
{\bf{Acoustic waveguide sensor}}. We used a Love mode SAW sensor obtained by coating a 0.5 mm thick ST-cut quartz substrate with a $1.2\ \mathrm{\mu m}$ thick layer of silicon dioxide by plasma enhanced chemical vapor deposition (Plasmalab 100, Oxford Plasma Technology, UK). The interdigital transducers (IDTs) were etched in a 200 nm thick aluminum layer sputtered on the substrate consisting of 100 pairs of split fingers with a periodicity of $40\ \mathrm{\mu m}$ and an acoustic aperture of $3.2\ \mathrm{mm}$. The acoustic filter had its central frequency at $123.5\ \mathrm{MHz}$. The sensing area between the transducers was covered with a $10\ \mathrm{nm}$ titanium, $50\ \mathrm{nm}$ gold coating that was electrically grounded to shield the acoustic waveguide. The sensor was packaged using a technique described elsewhere \cite{Francis04} that defined a $180$ to $200\ \mathrm{\mu l}$ open well above the sensing area of $D=4.7\ \mathrm{mm}$ length and isolated the sensing area from the electrode area in order to reduce electrical interactions between the IDTs and the liquid. The open well formed a static liquid cell in which the solution was manually injected using a micropipette. The SAW phase and insertion loss were continuously monitored in an open loop configuration with an HP4396A Network Analyzer and logged on a PC through the GPIB interface.\\

{\bf{Combined SAW/SPR setup.}} The setup has been described elsewhere \cite{Friedt04}. Briefly, a commercial Ibis II SPR instrument (IBIS Technologies BV, Hengelo, The Netherlands) based on a $670\ \mathrm{nm}$ laser diode has been modified in order to replace the usual gold-coated glass slide with the packaged Love mode SAW sensor.\\

{\bf{Temperature monitoring.}} A type K thermocouple (Chromel-Alumel) junction was made by wire-bonding two $100\ \mathrm{\mu m}$-diameter wires (Filotex, France) and was monitored by an AD595AD chip connected to an HP3478A multimeter. The temperature of the solution right above the sensing area was continuously monitored during the experiment and synchronously with the SAW and SPR measurements.\\

{\bf{Surface preparation.}} The device was cleaned in UV-O$_3$ for 30 minutes and followed by a 3 hours contact coating with octadecanethiol hydrophobic self-assembled monolayer obtained from Sigma-Aldrich.\\

{\bf{poly(N-isopropyl acrylamide)}} (PNIPAAm) from Polysciences Inc. (Warrington, PA, USA) was obtained with an average molecular weight $M_w=40\ \mathrm{kD}$ and a polydispersity index ($M_w/M_n$) of 2 as characterized by the manufacturer. A solution of 500 ppm polymer in deionized (DI) water (clean room grade, $18\ \mathrm{M\Omega .cm}$) was prepared, dialyzed to cut-off molecular weight of 12-14 kD (dialysis tubing DTV1200 from Medicell International Ltd, UK) over night and stored at $4\ \mathrm{^\circ C}$.\\

{\bf{Experimental method.}} The temperature dependence of the thiol coated SAW sensor was first calibrated in DI water by cycling from $20\ \mathrm{^\circ C}$ to $40\ \mathrm{^\circ C}$ (rising temperature rate: $85\pm 2\ \mathrm{K/h}$, decreasing rate: $49\pm 2\ \mathrm{K/h}$). After calibration, the 500 ppm polymer solution was injected while the temperature was set at $29\pm 1\ \mathrm{^\circ C}$. Following the procedure described by Plunkett {\it{et al.}} \cite{Plunkett203}, the sensing area was washed with twice the volume of DI water after 3 hours at a constant temperature of $29\pm 1\ \mathrm{^\circ C}$. A third volume of DI water was then injected for the measurement. The temperature was then cycled several times from $\sim 20\ \mathrm{^\circ C}$ to $\sim 40\ \mathrm{^\circ C}$ to monitor the influence of the temperature induced conformation changes of the polymer on the acoustic and optical signals.\\

\section{Results}

The experimental results of SAW and SPR data are given after subtraction of the temperature calibration in Fig.~\ref{fig:rawdata} that provides the thermal cycle for the SAW insertion loss I.L., the SAW phase variation $\Delta \phi$ and the SPR coupling angle variation $\Delta \theta$ for the thermal cycling during the calibration step and after the polymer adsorption. The temperature calibrations were fitted by polynomial interpolations and subtracted from the raw data with expressions as a function of the temperature $T$,  in $\mathrm{^\circ C}$, as following:
\begin{itemize}
\item SAW insertion loss $\mathrm{I.L.\  (dB)}=1.27\cdot10^{-4}T^3-1.29\cdot10^{-2}T^2+0.422T-25.8$;
\item SAW phase $\phi\ \mathrm{(^\circ)} =2.57T+45.1$ (corresponding to a temperature coefficient of frequency - TCF - of $32\ \mathrm{ppm/^\circ C}$);
\item SPR angle shift $\Delta \theta\ \mathrm{(m^\circ)} = -22.7T+1797$.
\end{itemize}
The observed average shift with temperature is compatible with the tabulated variation of bulk water refraction index \cite{CRC75} while the average shifts of the SAW data are following more complicated patterns that involve the effect of the transducers and the thermal shift of the acoustic propagation, not linearly correlated with temperature as in the SPR case.\\
The mechanical and optical constants used for the SAW and SPR models are given in Table~\ref{tb:matrdata} \cite{Brandrup66,CRC75,Palik97,FrancisUFFC04}. The evolution of the acoustic and optical data per thermal cycle is given in Fig.~\ref{fig:sawspr}. One temperature cycle starts for the lowest temperature and ends at the next temperature minimum. It shows the evolution of the ratio of the SAW attenuation to the SAW phase $\Delta A/\Delta \phi$ (left axis) and the SPR angle $\Delta \theta$ (right axis) for the first 4 cycles. The SAW/SPR data measured at 5 different points on each cycle ($25.0\ \mathrm{^\circ C}$ and $35.1\ \mathrm{^\circ C}$ during heating up, and $34.9\ \mathrm{^\circ C}$, $30.8\ \mathrm{^\circ C}$ and $25.0\ \mathrm{^\circ C}$ during cooling down) are reported in Table~\ref{tb:rawdata}; the point of $30.8\ \mathrm{^\circ C}$ with decreasing temperature corresponds to the peak value of $\Delta A/\Delta \phi$. The simulated values obtained with the methodology presented in the theoretical section are reported in Table~\ref{tb:thickness}. The evolution of the polymer weight concentration and the viscosity of the PNIPAAm-water layer is plotted in Figs.~\ref{fig:sprevolve} and \ref{fig:viscothick}, respectively, as a function of the layer thickness.

\section{Discussion}

The ratio of the SAW attenuation shift to the phase angle shift is a good indicator of the variation of viscosity or thickness of the film and of conformational changes that occur in the film, this analysis can be extended to films evolving at constant temperature such as during immunoassay reactions. Taken separately, the increase of the amplitude $\Delta A$ and the phase $\Delta \phi$ indicate that the film is getting more viscous and thicker, while their ratio determines which factor is predominant. As given in Fig.~\ref{fig:hoverdelta}, a proportionality exists between this ratio and $h\sqrt{\rho_C}/\sqrt{\eta_C}$; consequently, if density variations are small in comparison to fluctuations of the film thickness and viscosity, an increasing ratio denotes a decrease of the viscosity or a thickening of the layer while a decreasing ratio indicates the opposite, i.e. an increasing viscosity or a thinning layer.\\

As the film thickness gets over the penetration depth, the SAW signal ratio saturates and the acoustic wave sees the deposited film as a semi-infinite bulk fluid. In such case, viscosity variations are probed through the individual variation of phase or insertion loss of the device. Hence, measuring the SAW signal ratio for viscous fluids is meaningless since its value is constant and equals a saturation value specific to each particular SAW sensor. In the same order of idea, a direct comparison of the SAW phase shift with the one obtained from a standard calibration of the sensor with a bulk viscous solution is not adequately fitted to the actual event that takes place on the sensing surface during the segregation of a soft layer: at each time that the film is smaller than the penetration depth, the film thickness plays a crucial role and must be determined independently to estimate correctly the relevant characteristics of the adsorbed film, as presently for the PNIPAAm case.\\

PNIPAAm presents a conformational transition temperature, also called Lower Critical Solution Temperature (LCST), that identifies a clear separation of the physical characteristics of the polymer as seen qualitatively by acoustic and optical techniques. The transition corresponds to a switch from hydrophilic to hydrophobic because of the unshielded isopropyl group \cite{Schild92,Lin99,Katsumoto02}. The LCST noticed by the abrupt change of SAW and SPR signals as seen in Fig.~\ref{fig:sawspr} is located at $30.8\pm 0.3\ \mathrm{^\circ C}$ for both rising up and cooling down. The transition is accompanied by a modification of the solubility of the polymer in solution and therefore by a gradient of concentration from the surface to the volume. Through this paper, it is assumed that the composite layer is PNIPAAm with water as solvent and a step gradient in concentrations defines a clear interface between the adsorbed layer and the bulk fluid. Practically, it is more likely that the layer-fluid interface displays a smoother gradient due to dangling polymer strands and polymer diffusion. According to the same principle, surface roughness effects known to influence the acoustic signal are disregarded at the moment as the surface roughness has been measured with a value in the order of 3-5 $\mathrm{nm_{pp}}$ \cite{Friedt03}, which is much smaller than the acoustic penetration depth and the film thickness. These specific effects, gradient of concentration and surface roughness, were not considered since the presented methodology is not complete enough to address such interfaces yet. To that end, further assumptions on the concentration distribution as a function of the depth are needed and must be implemented in the model as, for instance, a multilayered composite layer with varying mechanical characteristics following a diffusion law. This type of implementation requires additional theoretical and experimental work to establish and measure concentration gradients and surface roughness in order to determine their exact incidence on the quantitative results presented here. Such developments are far beyond the scope of this paper.\\

Qualitatively and quantitatively, several variations are detected for each thermal cycle, with consistency from cycle to cycle. Variations occur in all observed parameters - density, viscosity and film thickness - through the cycle. The discussion is facilitated by distinguishing 3 regions per thermal cycle: (I) from  $25\ \mathrm{^\circ C}$ to the LCST; (II) from LCST with temperature increasing to LCST with temperature decreasing; (III) from LCST to $25\ \mathrm{^\circ C}$. Fig.~\ref{fig:lcst} is a pictorial representation of the assumed mechanisms probed by the combined SAW/SPR technique and discussed here after.\\ In region (I), a 4 wt \% PNIPAAm hydrophilic layer absorbed on the surface is at equilibrium with the bulk solution. As the LCST is reached, the conformational transition of the polymer results, in region (II), in a radical and direct modification of the concentration that jumps almost immediately to a denser layer (12 wt \% ), thus a high viscosity resulting from stronger chains interactions. During the same time, the polymer in solution has a lower solubility and is segregating to the layer that continuously grows till a novel equilibrium is reached between surface and bulk concentrations. Meanwhile and as seen in Fig.~\ref{fig:viscothick}, the viscosity increases as more chains interactions are taking place. This explanation is correlated with the SPR signal that presents a sudden variation after the LCST is reached, reflecting the conformational change of the layer. On the other hand, the abrupt transition is seen in the SAW ratio that presents a slow ramp up in region (I) followed by a slight decrease in region (II). As mentionned before, a decreasing ratio $\Delta A/\Delta \phi$ indicates that the viscosity increases or the thickness decreases, the situation that prevails here is an increasing viscosity.\\

The situation that occurs during cooling is different as observed at the transition between regions (II) and (III). As the temperature reaches again the LCST, the polymer is back to a hydrophilic behavior and start to swell in water. Because of the swelling, the layer thickness slightly increases but the PNIPAAm concentration and the viscosity are dropping fast. The effect is seen in the SAW signal ratio that is ramping up since a decrease of viscosity increases the ratio. Immediately after, in the early stage of region (III), the layer is effectively swelling and gets thinner while the viscosity remains constant, therefore the ratio is coming back to the initial value it presented in region (I). The estimation of the viscosity of organic monolayers is a specific feature displayed by the combined technique while few methods are able to tackle this type of measurement \cite{Barentin00}, so an exact comparison of our results with literature values is not directly possible although the estimated values determined by simulations are reasonable and of the same order of magnitude as the ones reported by Milewska {\it{et al.}} \cite{Milewska03} for bulk solutions of PNIPAAm although they characterized PNIPAAm solutions with a larger molecular weight of 525 kD.\\

The asymmetry seen in the intensity of the SAW signal ratio peaks at LCST is possibly linked to the large difference in rising and decreasing temperature rates. It has been reported by Boutris {\it{et al.}}\cite{Boutris97} from optical density measurements that the scanning rate influences the temperature transition because the system is not at equilibrium for large scanning rates; in our measurement, the scanning rate is moderate and the LCST is seen at the same temperature for both directions of the thermal cycling, indicating that the system is at equilibrium at any moment of the cycle otherwise we would observe a shift in the location of the LCST as a function of the temperature direction. Boutris {\it{et al.}} equally reported that the dissolution of the PNIPAAm during cooling is a much slower process for an identical temperature scanning rate than the abrupt transition noticed during the heating up, which might account for the intensity difference of the SAW signal ratio seen upon heating up and cooling down. Finally, as the number of cycles increases, a part of the polymer is diffusing away from the surface and will not re-enter the surface-volume equilibrium process of the following thermal cycle, explaining for the cycle degradation.\\

The temperature change might also cause modification to the interpretation of the data and are important for the calibration. The SPR sees the refractive index modification quite clearly and is linearly dependent upon the change with a given slope. The SAW situation differs from the SPR: with temperature, the water viscosity drops from $\simeq$1 cP to 0.6 cP  from $20\ \mathrm{^\circ C}$ to $35\ \mathrm{^\circ C}$ as tabulated \cite{CRC75}, corresponding theoretically to lower value of the insertion loss and a phase shift but in the same time the propagation characteristics of the Love mode device are modified by the temperature shift, the net result is the 3$^{rd}$ order polynomial expression for the temperature on the insertion loss of the device while it is included in a linear relation for the phase. The effects compensate for each other.\\

Our observations are in agreement with the results reported by Plunkett {\it{et al.}} of a dense wetting layer that grows in thickness above the LCST. Our method provides an additional quantitative evaluation of the layer viscosity below and above the transition temperature, and quantitative information determining the amount of solvent and the thickness of the adsorbed soft layer.\\

\section{Conclusions}

The combination of surface acoustic wave and surface plasmon resonance measurements delivers useful information to estimate the in situ properties of surface adsorbed layers. A general method of data extraction has been presented that can be applied to layers adsorbed from liquid environments, such as in biosensors. The experimental ratio of amplitude shift to phase shift of the SAW signal correlated to simulations is an indicator of a rigid or a viscous interaction between the layer and the acoustic wave. A soft layer is characterized by a viscous behavior that results in a high damping of the acoustic wave, thus to a high ratio of the attenuation to the phase shift. The ratio is not a sufficient criterion to discriminate between soft and rigid layers since for very thin adsorbed layers, the two possible behaviors have a similar effect on the acoustic signal. Without data analysis, the experimental ratio is an indicator of the evolution of the thickness or of the viscosity of the layer: an increasing ratio is related to a growing film or to a viscosity decrease, while a decreasing ratio is related to a thinning film or to a viscosity increase.\\

With simulations of an equivalent viscoelastic transmission lines model, SAW and SPR data are used to assess quantitatively the density, the viscosity and the thickness of soft layers. The potential of the method has been demonstrated with the example of PNIPAAm, a thermally sensitive polymer. The record of SAW and SPR data during a thermal cycle clearly shows a conformational transition at the LCST point measured at 30.8 $\mathrm{^\circ C}$. The LCST transition is seen in the SAW signal ratio by a peak value and in the same time by an abrupt  variation of the SPR signal. Graphs showing the evolution of the PNIPAAm weight concentration and viscosity versus the layer thickness were built that indicate first a thickening of the layer followed by an increase of viscosity then the collapse of the layer. We also conclude from this study that the analysis of the adsorbed layer of PNIPAAm can be considered under the assumption of a soft layer as determined by the high value of the experimental SAW signal ratio $\Delta A/\Delta \phi$ ratio which is in agreement with a viscous interaction between the layer and the acoustic wave. Simulation results validate the soft layer hypothesis as seen by a high water content in the layer (between 88 wt \% and 96 wt \%).

\section{Acknowledgments}

The authors wish to thank R. Giust (LOPMD, Besan\c con, France) for kindly providing SPR simulation routines. L. Francis would like to acknowledge the Fonds pour la Formation \`a la
Recherche dans l'Industrie et dans l'Agriculture (F.R.I.A., Belgium) for financial support.

\newpage
Fig.~\ref{fig:pnipaam}. Schematic of the PNIPAAm structure for a degree of polymerization $n$.\\

Fig.~\ref{fig:sawspr1}. Schematic cross-section of the combined SAW/SPR setup: a SAW device is attached with optical index matching oil to a prism for the coupling of a LASER light with the gold coated surface of the acoustic device to generate the surface plasmon resonance. The arrows indicate the propagation of the acoustic wave (dashed line) and the collimated LASER beam (dot-dashed line).\\

Fig.~\ref{fig:rlc_model}. Differential element of the viscoelastic mechanical transmission line model: the shear stress ($T$) and the shear wave particle velocity ($v$) propagate through a symmetric differential element of length $dr$ made of an inductance $L$ shunted by the series connection of a capacitance $C$ and a conductance $G$.\\

Fig.~\ref{fig:tlm_mod12}. Schematic representation of the physical layered structure of the Love mode SAW sensor. A guided acoustic shear-acoustic wave propagates in the $x$ direction in a stack composed of the semi-infinite piezoelectric substrate (S), a guiding layer (G) of thickness $t$, the composite layer (C) of thickness $h$ that segregates from a semi-infinite fluid (F). The sensing area of the device is at the interface between the guiding layer and the adsorbed composite layer. The plane T is a reference plane used to determine the propagation characteristics of the guided acoustic wave. For clarity, only one guiding layer is represented although the practical device is composed of several guiding layers.\\

Fig.~\ref{fig:tlm_mod11}. Equivalent transmission line model of the physical structure in the direction of resonance ($z$) given in Fig.~\ref{fig:tlm_mod12}. The equivalent transmission lines of each part of the device implement the model of Fig.~\ref{fig:rlc_model} and are schematically represented by stub lines with the indication of the wavenumber $k_{jz}$ and the characteristic impedance $Z_{jz}$. The lines are connected in series to satisfy the boundary conditions and the two semi-infinite layers are loaded with a matching impedance. Resonance solutions of this network are solution for a guided Love mode.\\

Fig.~\ref{fig:hoverdelta}. Simulated values for the ratio of the SAW attenuation shift to the SAW phase shift $\Delta A/\Delta \phi$ as a function of the composite film thickness normalized to the penetration depth in this film $\delta = \sqrt{2\eta_C/\left(\rho_C\omega \right)}$.\\

Fig.~\ref{fig:rhoeta_saw1}. Plot for a given film thickness of the values of normalized density and normalized viscosity for a given and constant value of the SAW phase shift $\Delta \phi$ (solid line) and a constant value of the SAW insertion loss shift $\Delta A$ (dashed line). The values of density are normalized to obtain the direct correlation with SPR data. The intersection point of the two curves gives for that thickness the density and viscosity of the film. The procedure must be repeated for all possible values of the film thickness.\\

Fig.~\ref{fig:rawdata}. Raw SAW and SPR data temperature calibration and temperature cycling effect as a function of the temperature for 5 cycles after removal of the temperature effect. The arrows indicate the direction of the temperature cycling. Top graph: SAW insertion loss, middle graph: SAW phase, and bottom graph: SPR angle.\\

Fig.~\ref{fig:sawspr}. Evolution of the SAW attenuation shift to phase shift ratio (dots, left axis) and SPR shift (squares, right axis) for the 4 first temperature cycles.\\

Fig.~\ref{fig:lcst}. Pictorial representation of the mechanisms probed by the SAW/SPR. Below LCST, the layer contains a large amount of water and because of the solubility of the physisorbed PNIPAAm, there is an equilibrium between surface and bulk concentrations. As the temperature increases above LCST, the conformation changes and results in an hydrophobic material collapsing on the surface, in addition to a segregation: as a result, a denser and thicker layer builds above the surface. As diffusion drives part of the material away from the surface, SAW and SPR signals are getting different as the number of thermal cycles increases.\\

Fig.~\ref{fig:sprevolve}. Summarizing graph of the evolution of the estimated PNIPAAm weight concentration in the adsorbed layer and the thickness for the 5 probed temperatures. The background lines are the measured SPR angle shifts.\\

Fig.~\ref{fig:viscothick}. Summarizing graph of the evolution of the estimated viscosity of the adsorbed layer and the thickness for the 5 probed temperatures.\\

Table~\ref{tb:matrdata}. Mechanical and optical constants used for the SAW and SPR models for the materials entering the experimental set-up and for a device operating at a frequency of 123.5 MHz.\\

Table~\ref{tb:rawdata}. Experimental mean values observed during the 4 first temperature cycles probed at 5 different points for the thermocouple, the SAW and the SPR. The arrows indicate the direction of the temperature cycle.\\

Table~\ref{tb:thickness}. Evaluation of the density $\rho$, viscosity $\eta$, thickness $h$, penetration depth in the film $\delta$ simulated with the results indicated in Table~\ref{tb:rawdata}. The simulated thickness to penetration depth ratio $h/\delta$ and the experimental SAW amplitude shift to SAW phase shift $\Delta A/\Delta \phi$ are given for a comparison with Fig.~\ref{fig:hoverdelta}.\\

%
%

\newpage
\begin{figure}
    \centering
    \caption{Schematic of the PNIPAAm structure for a degree of polymerization $n$.}\label{fig:pnipaam}
\end{figure}

\newpage
\begin{center}
\begin{figure}
        \includegraphics[width=\columnwidth]{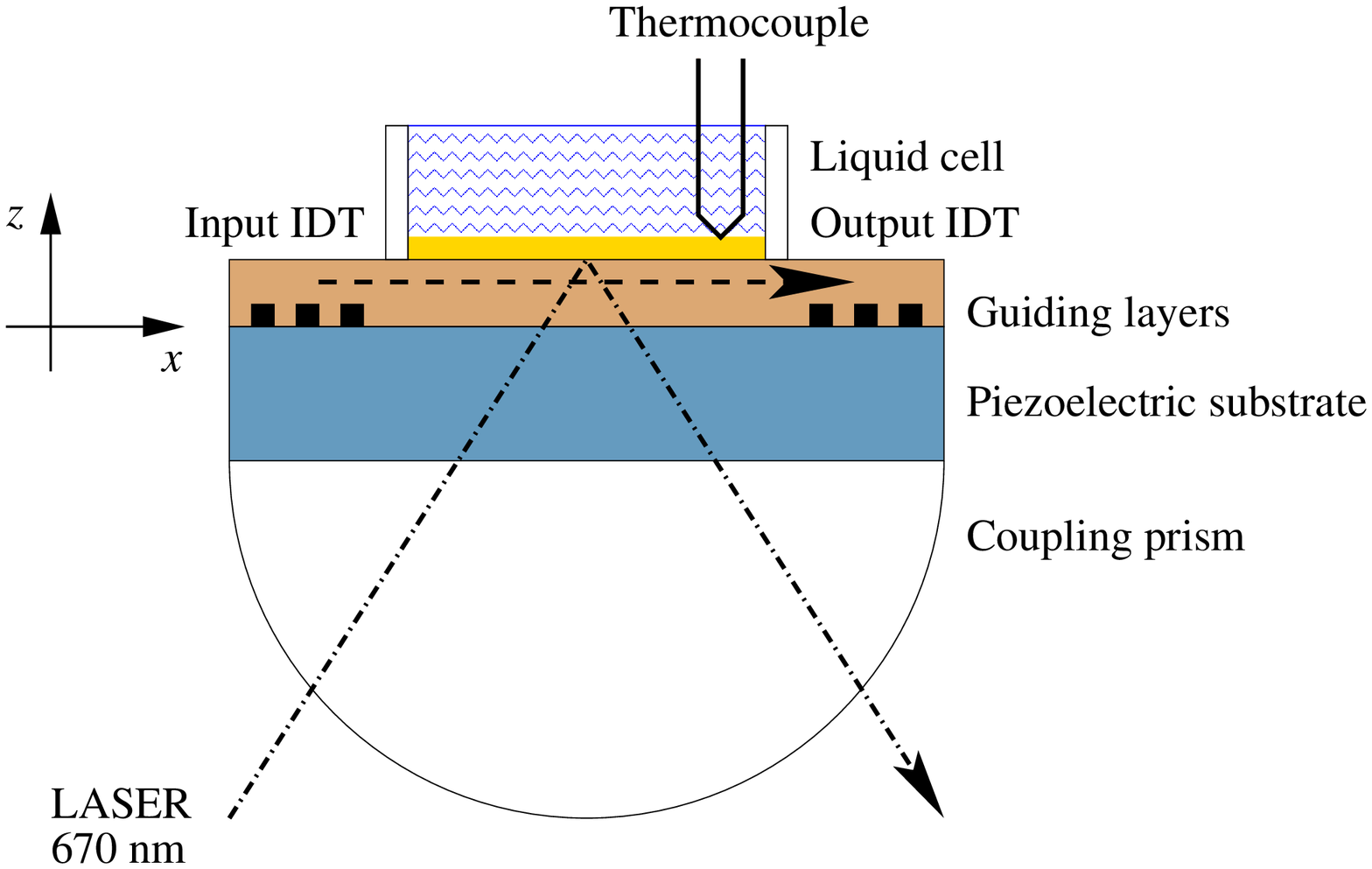}
    \caption{} \label{fig:sawspr1}
\end{figure}
\end{center}

\newpage
\begin{center}
\begin{figure}
        \includegraphics[width=\columnwidth]{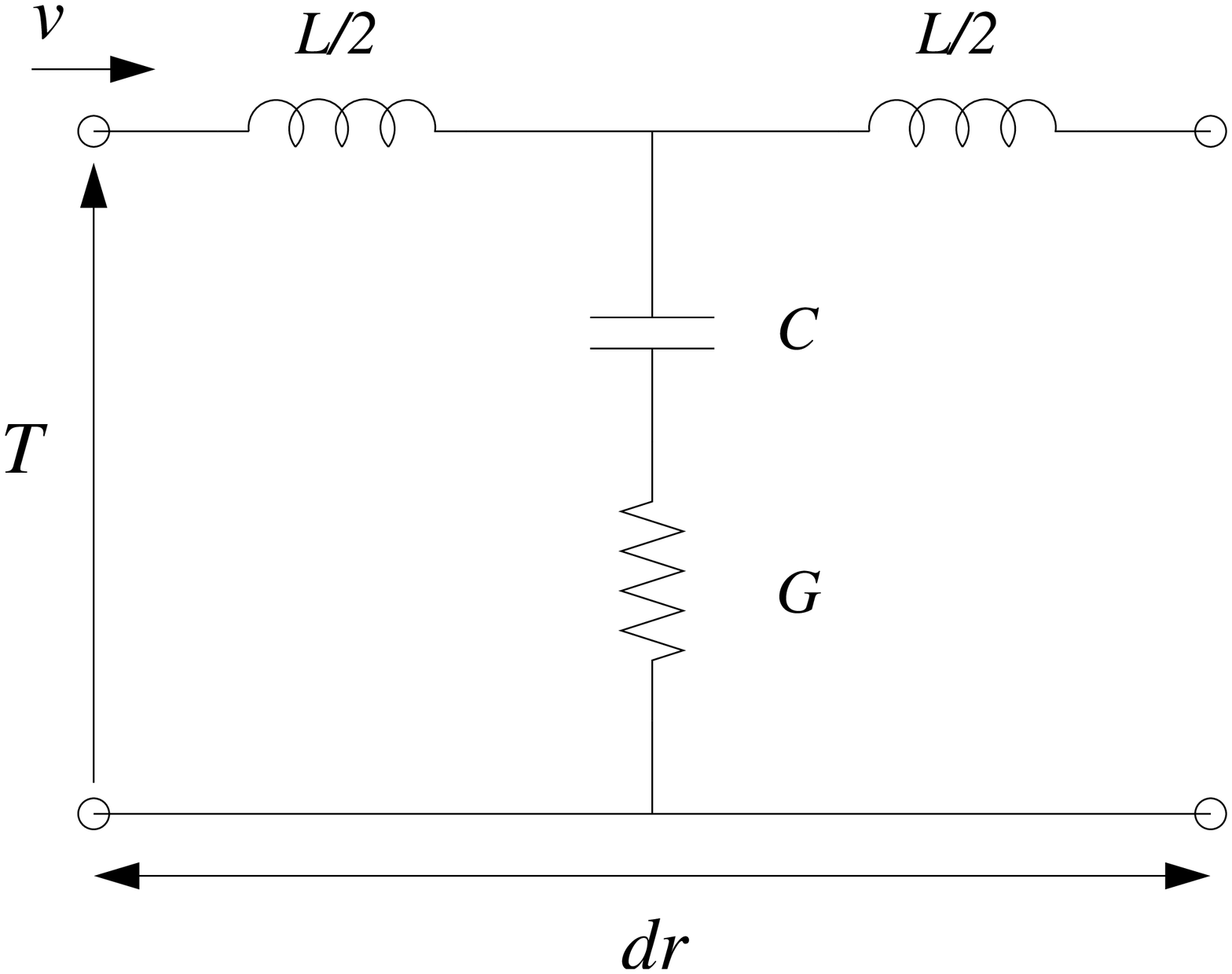}
    \caption{} \label{fig:rlc_model}
\end{figure}
\end{center}

\newpage
\begin{center}
\begin{figure}
        \includegraphics[width=\columnwidth]{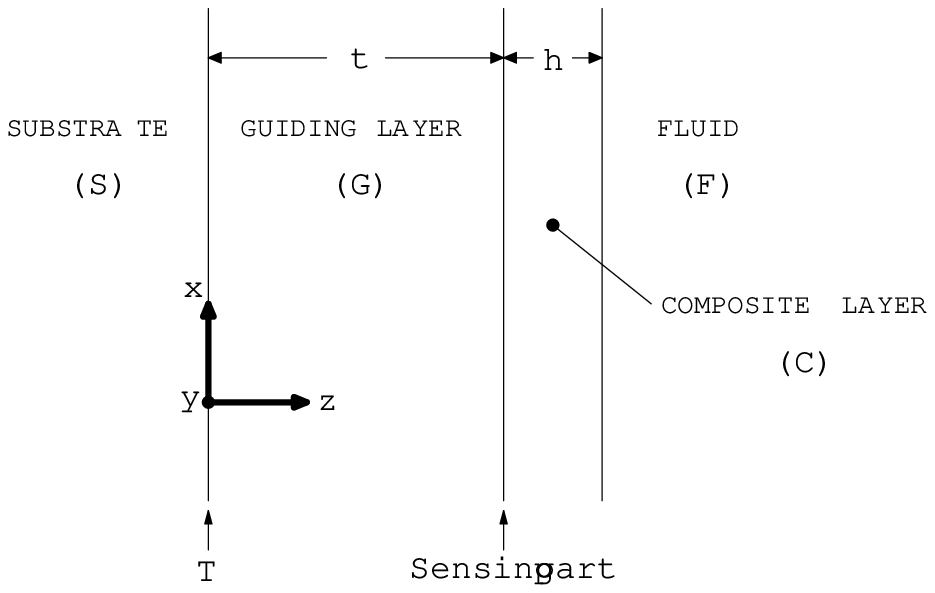}
    \caption{} \label{fig:tlm_mod12}
\end{figure}
\end{center}

\newpage
\begin{center}
\begin{figure}
        \includegraphics[width=\columnwidth]{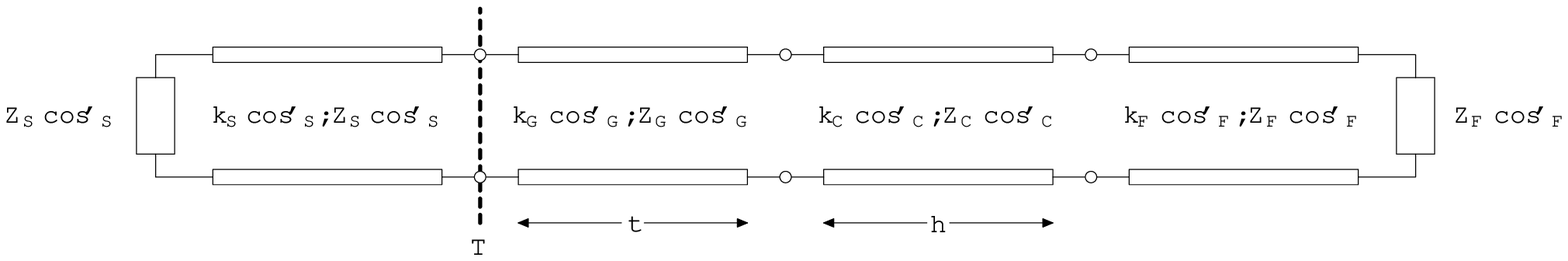}
    \caption{} \label{fig:tlm_mod11}
\end{figure}
\end{center}

\newpage
\begin{center}
\begin{figure}
        \includegraphics[width=\columnwidth]{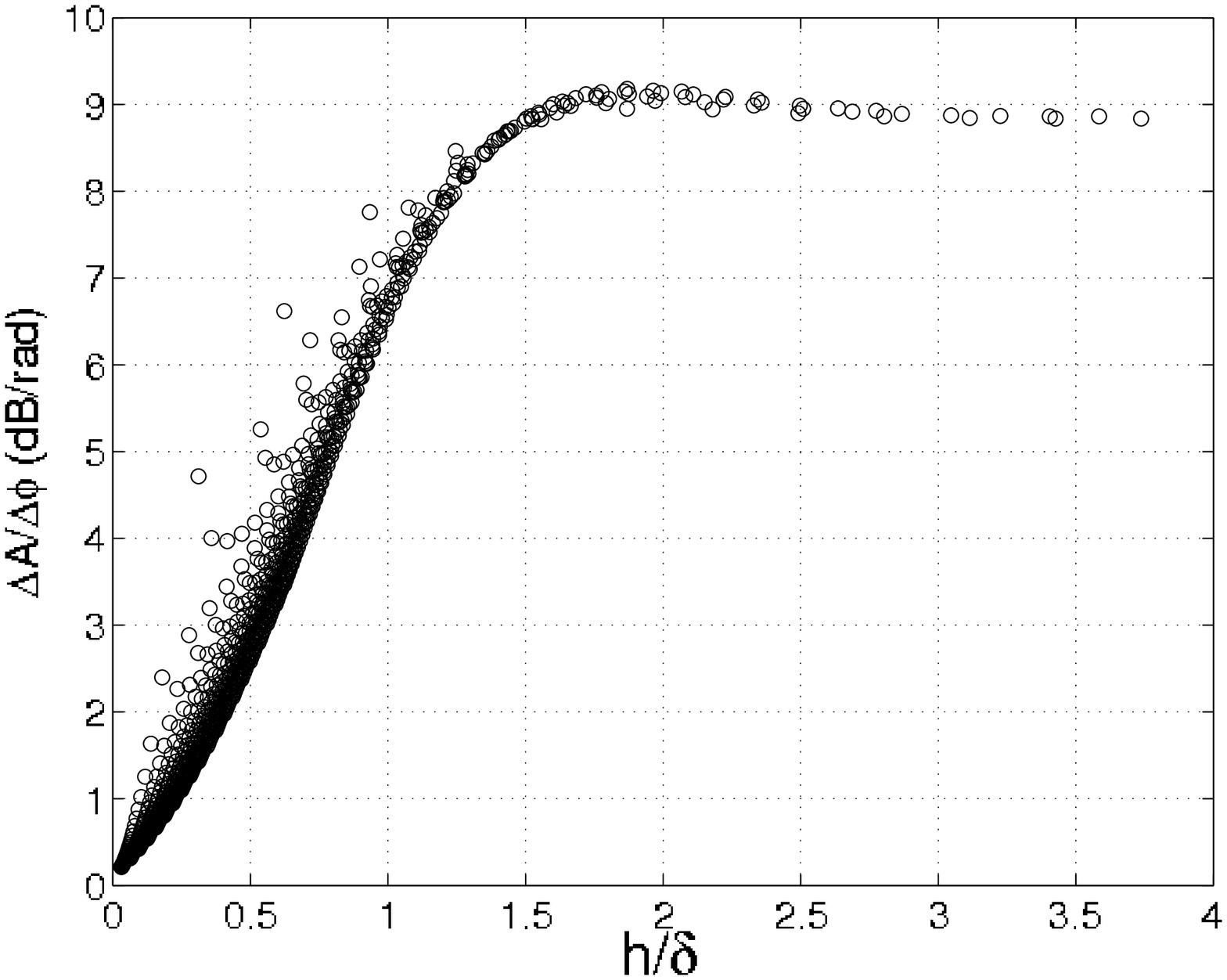}
    \caption{} \label{fig:hoverdelta}
\end{figure}
\end{center}

\newpage
\begin{center}
\begin{figure}
        \includegraphics[width=\columnwidth]{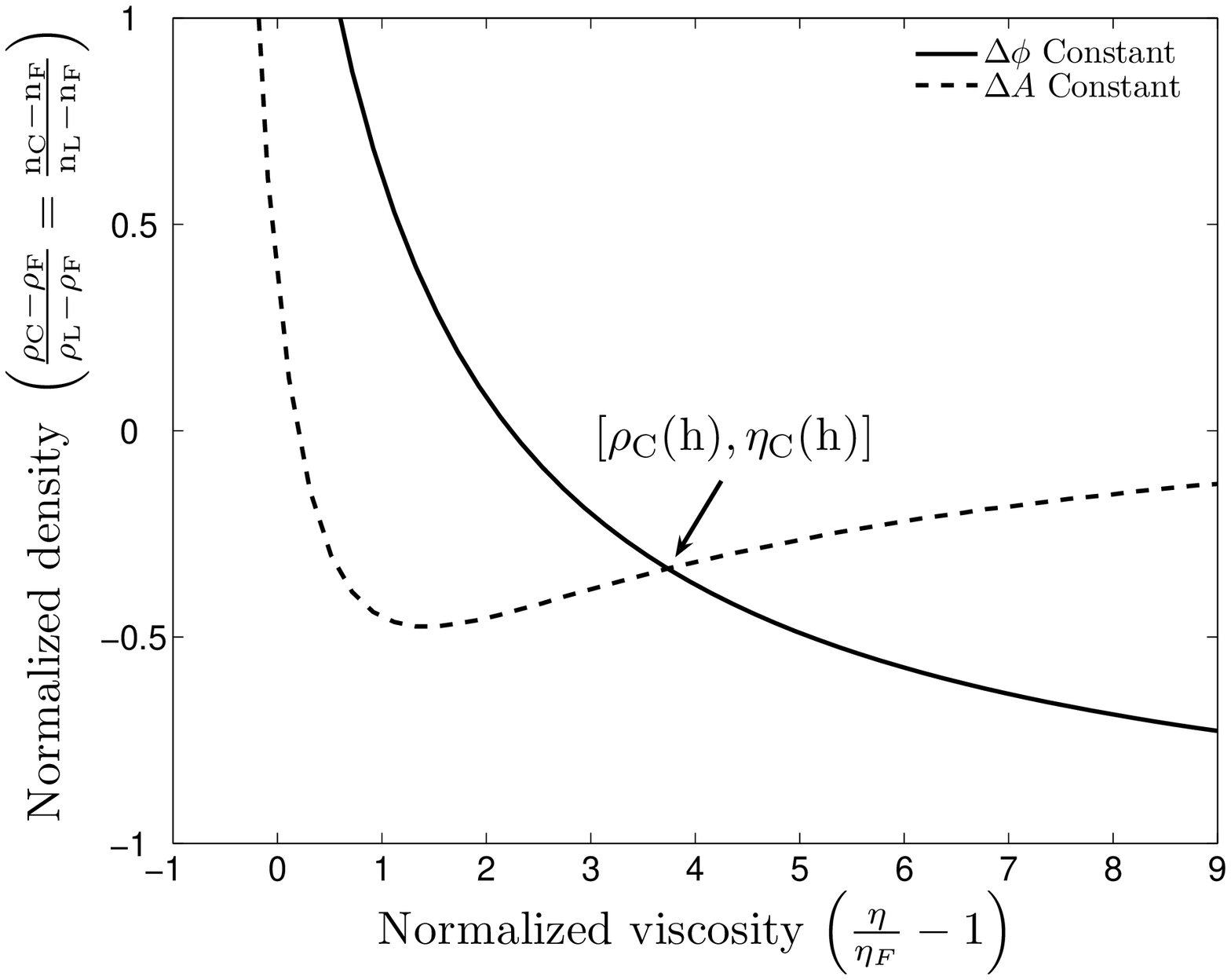}
    \caption{} \label{fig:rhoeta_saw1}
\end{figure}
\end{center}

\begin{center}
\begin{figure}
        \includegraphics[width=\columnwidth]{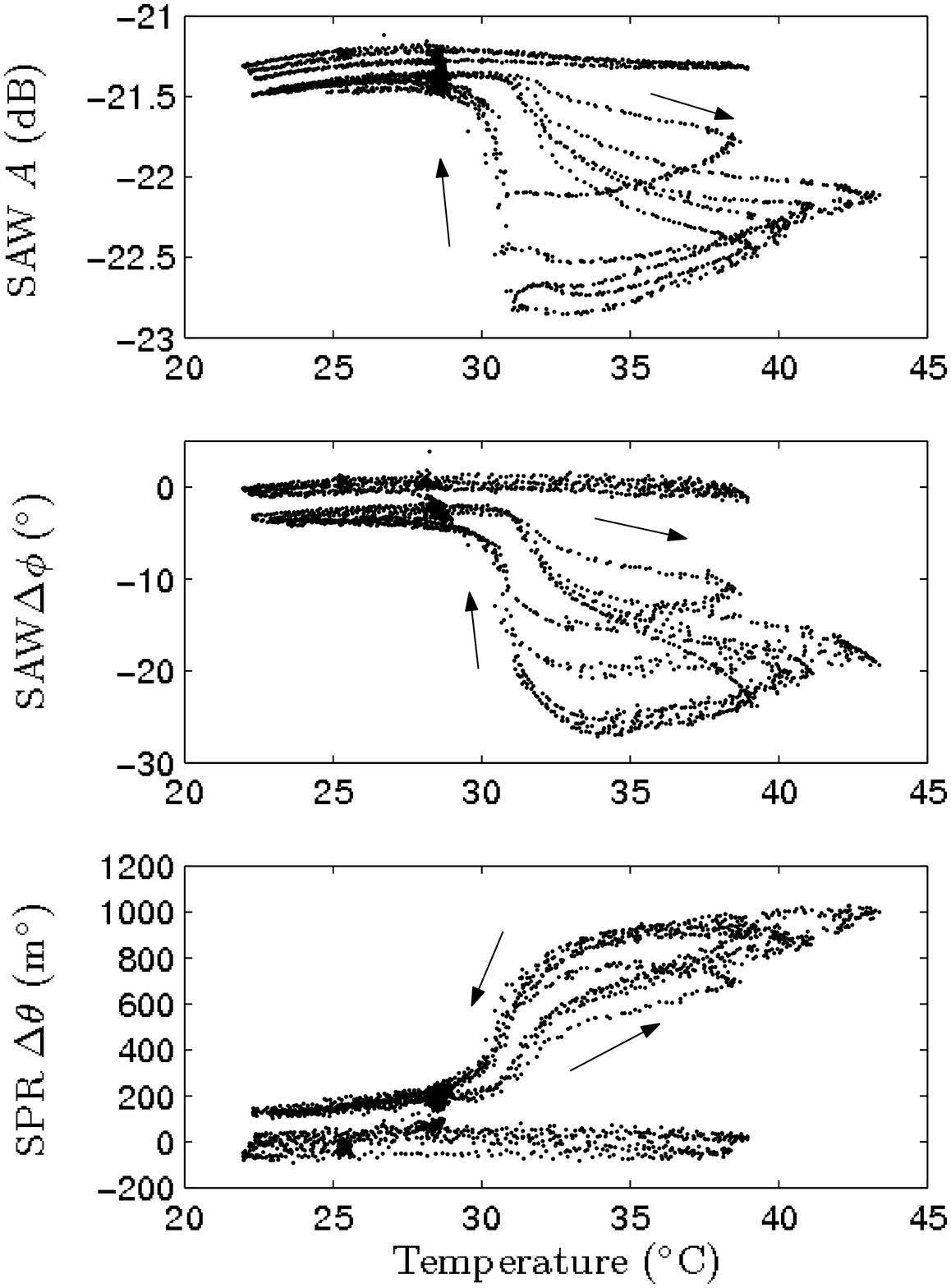}
    \caption{} \label{fig:rawdata}
\end{figure}
\end{center}

\newpage
\begin{center}
\begin{figure}
        \includegraphics[width=\columnwidth]{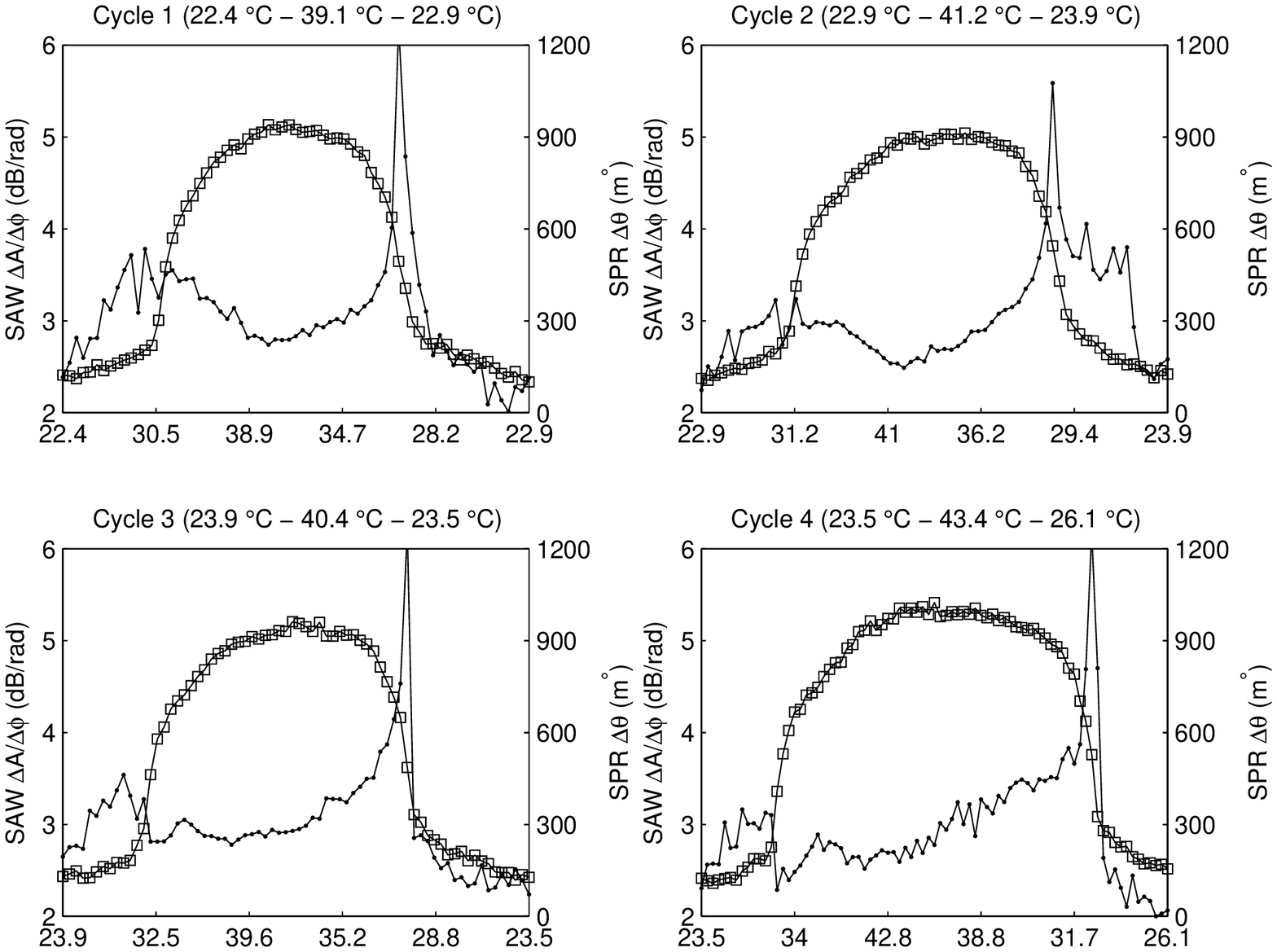}
    \caption{} \label{fig:sawspr}
\end{figure}
\end{center}

\newpage
\begin{center}
\begin{figure}
        \includegraphics[width=\columnwidth]{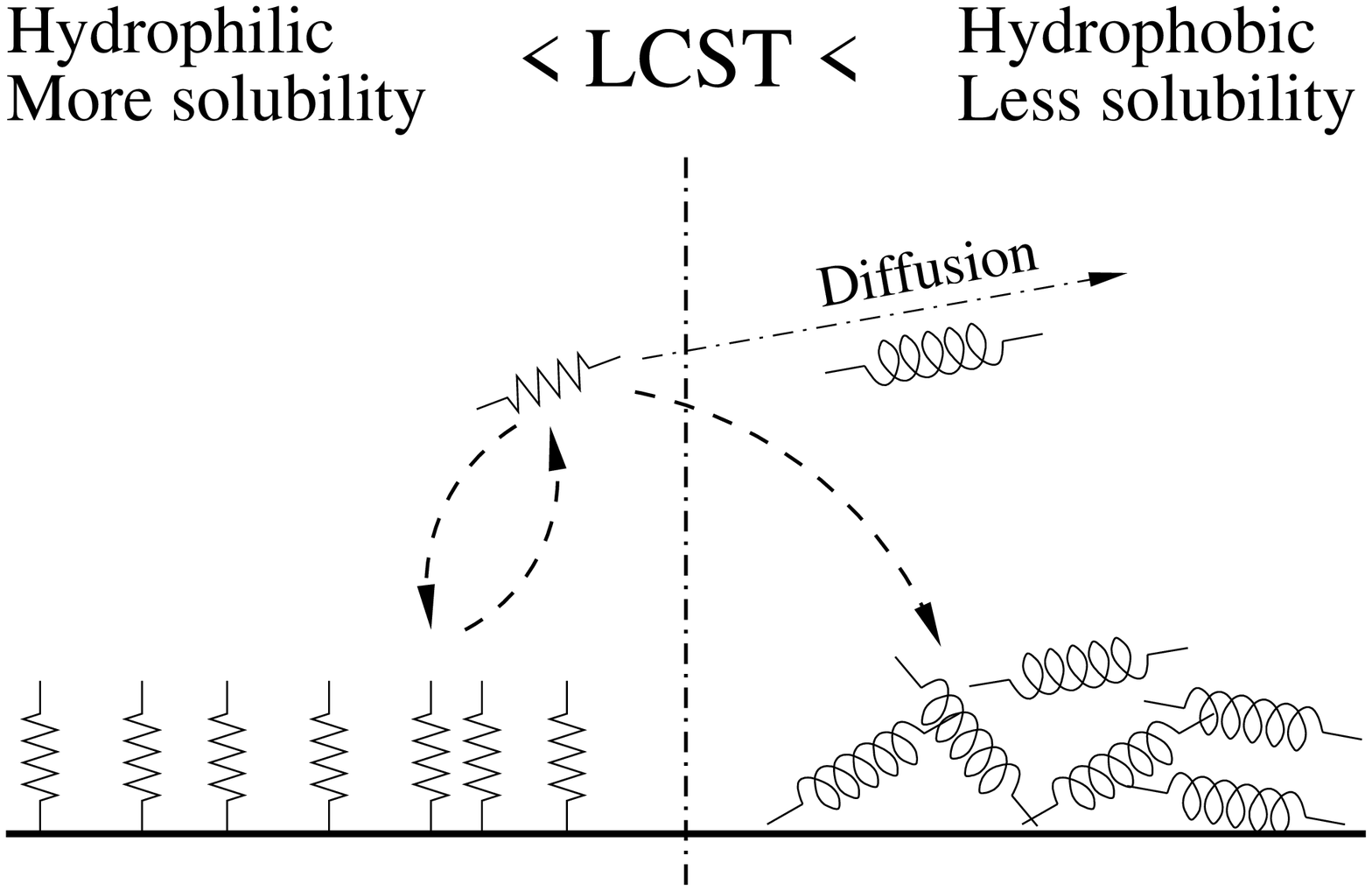}
    \caption{} \label{fig:lcst}
\end{figure}
\end{center}

\newpage
\begin{center}
\begin{figure}
        \includegraphics[width=\columnwidth]{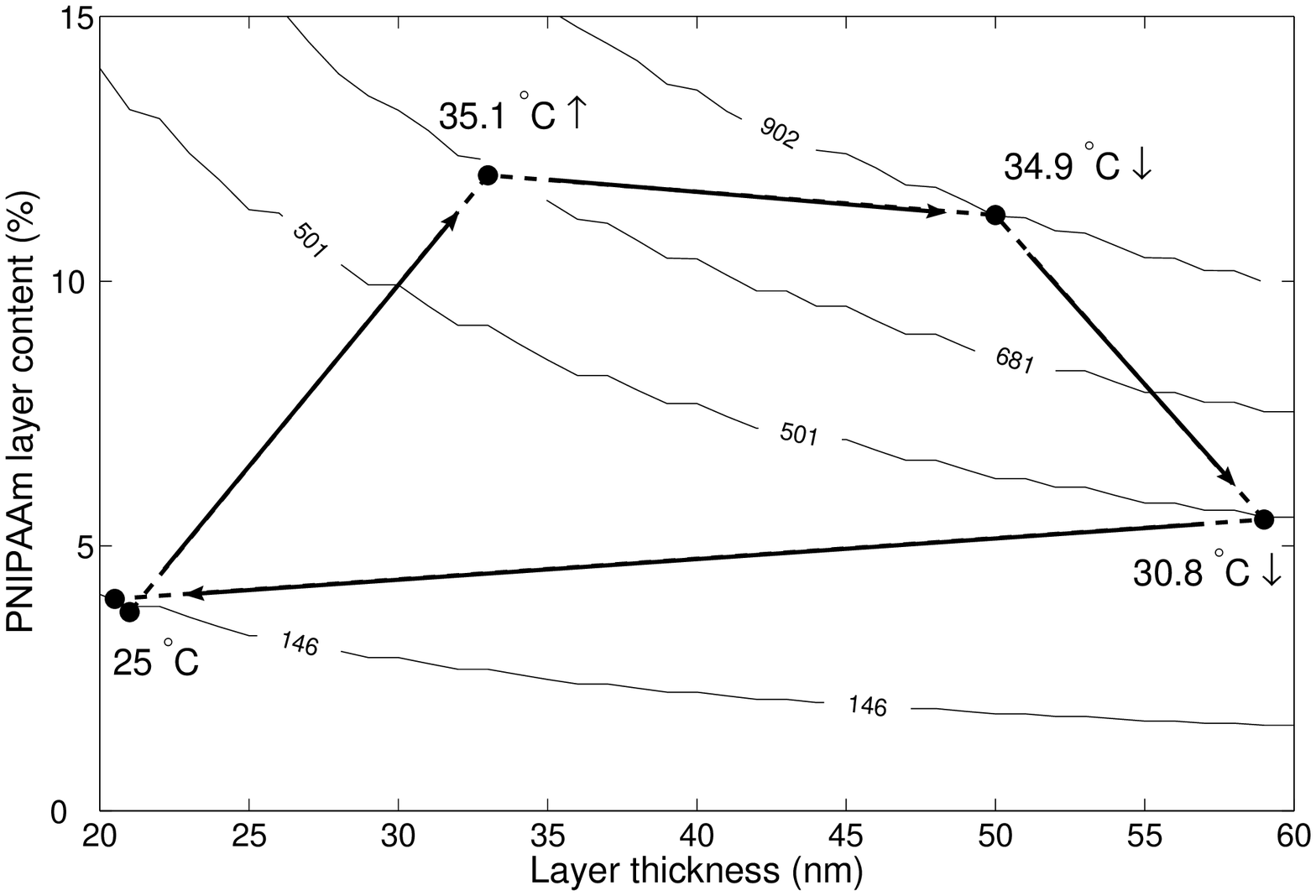}
    \caption{} \label{fig:sprevolve}
\end{figure}
\end{center}

\newpage
\begin{center}
\begin{figure}
        \includegraphics[width=\columnwidth]{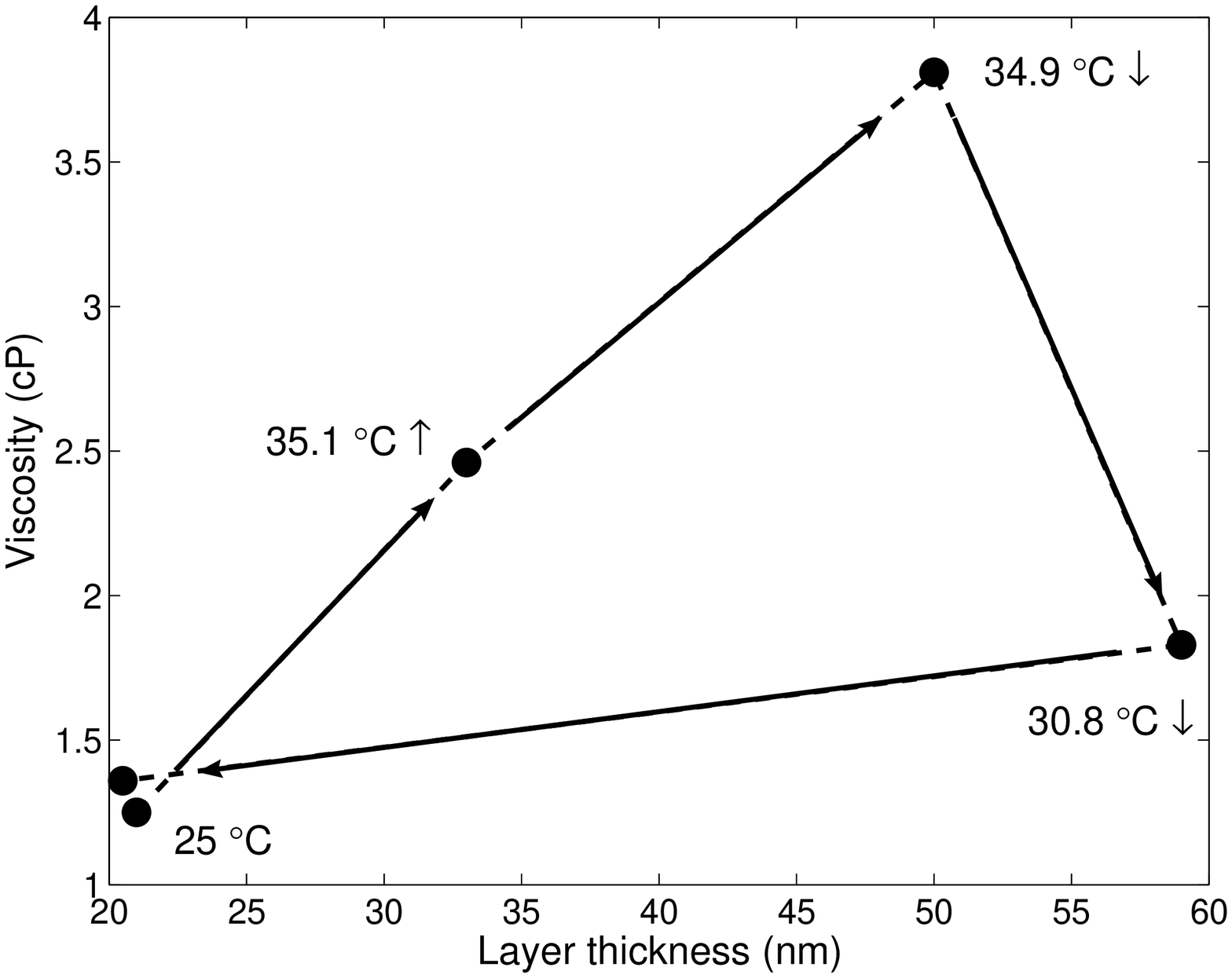}
    \caption{} \label{fig:viscothick}
\end{figure}
\end{center}

\newpage
\begin{table}[h!tb]
{\small \begin{tabular}{c|c|c|c|c|c} \hline

{\bf{Material}}& {\bf{Density}} & {\bf{Shear stiffness}} & {\bf Viscosity} & {\bf{Ac. Velocity}} & {\bf{Refractive}} \\
& $\rho\ (\mathrm{g/cm^3})$ & $\mu\ (\mathrm{GPa})$ & $\eta\ (\mathrm{cP})$ & $V\ (\mathrm{m/s})$ & {\bf index} $n$\\ \hline \hline
Quartz & $2.648$ & $66.3$ & 0 & $5004$ & $1.518$ \\
$\mathrm{SiO_2}$ & $2.61$ & $8.31$ & 0 & $1784$ & $1.5$ \\
Ti & $4.51$ & $46.7$ & 0 & $4230$ & $2.76+3.84\mathrm{i}$ \\
Au & $19.3$ & $28.5$ & 0 & $1215$ & $0.14+3.697\mathrm{i}$ \\
Water & $1$ & 0 & $1$ & $19.7+19.7\mathrm{i}$ & $1.33$ \\ \hline
PNIPAAm & $1.4$ & & & & $1.47$ \\ \hline\end{tabular}
}
\caption{}
\label{tb:matrdata}
\end{table}

\newpage
\begin{table}[h!tb]
{\small \begin{tabular}{c|c|c|c} \hline

{\bf{Thermocouple}}& \multicolumn{2}{|c|}{{\bf{SAW}}} & {\bf{SPR}}\\

Temp. ($\mathrm{^\circ C}$) & $\Delta A$ (dB) & $\Delta \phi$ ($\mathrm{^\circ}$)&  $\Delta \theta$ ($\mathrm{m^\circ}$)\\ \hline \hline

$25.0\pm0.1$ $\searrow$& $-0.17\pm0.04$ & $-3.60\pm0.38$ &   $146\pm17$ \\

$25.0\pm0.1$ $\nearrow$& $-0.14\pm0.02$ & $-2.72\pm0.74$ &   $142\pm13$ \\

\hline

$30.8\pm0.3$ $\searrow$& $-1.05\pm0.43$ & $-10.0\pm2.4$ &  $501\pm21$ \\

\hline

$34.9\pm0.3$ $\searrow$ & $-1.35\pm0.22$ & $-24.0\pm5.2$ &   $902\pm52$ \\

$35.1\pm0.1$ $\nearrow$& $-0.66\pm0.45$ & $-13.2\pm5.4$ &   $681\pm127$ \\

\hline\end{tabular}
}
\caption{}
\label{tb:rawdata}
\end{table}

\newpage
\begin{table}[h!tb]
{\small \begin{tabular}{c|c|c|c|c||c|c} \hline
{\bf{Temp.}} & ${\mathbf{\rho_C}}$ & ${\mathbf{\eta_C}}$ & ${\mathbf{h}}$ & ${\mathbf{\delta}}$ & ${\mathbf{h/\delta}}$ & $\Delta A/\Delta \phi$ \\
($\mathrm{^\circ C}$) & ($\mathrm{g/cm^3}$) & (cP)& (nm) & (nm) & & (dB/rad)\\
\hline \hline
$25.0$ $\searrow$ & $1.016$ & $1.36$ & $20.5\pm 1.3$ & $59$ & $0.35$ &$2.6\pm0.6$ \\
$25.0$ $\nearrow$ & $1.015$ & $1.25$ & $21\pm 1.3$ & $56$ & $0.37$ &$3.0\pm0.5$ \\ \hline
$30.8$ $\searrow$ & $1.022$ & $1.83$ & $59\pm 3.6$ & $68$ & $0.87$ &$5.9\pm1.3$ \\ \hline
$34.9$ $\searrow$ & $1.045$ & $3.81$ & $50\pm 5.0$ & $97$ & $0.51$ &$3.2\pm0.3$ \\
$35.1$ $\nearrow$ & $1.048$ & $2.46$ & $33\pm 2.0$ & $78$ & $0.42$ &$2.8\pm0.9$ \\
\hline\end{tabular}
}
\caption{}
\label{tb:thickness}
\end{table}


\end{document}